\documentclass[12pt]{article}
\def\be{\begin{equation}}\def\ee{\end{equation}}

 \makeatletter\ifcase\@ptsize\font\teneufm=eufm10
\font\seveneufm=eufm7\font\fiveeufm=eufm5
\font\teneusm=eusm10\font\seveneusm=eusm7
\font\fiveeusm=eusm5\or\font\teneufm=eufm10 scaled
\magstephalf\font\seveneufm=eufm7\font\fiveeufm=eufm5
\font\teneusm=eusm10 scaled\magstephalf
\font\seveneusm=eusm7\font\fiveeusm=eusm5\or \font\teneufm=eufm10
scaled\magstep1\font\seveneufm=eufm7
\font\fiveeufm=eufm5\font\teneusm=eusm10 scaled\magstep1
\font\seveneusm=eusm7\font\fiveeusm=eusm5\fi

\newfam\eufmfam\newfam\eusmfam\textfont\eufmfam=\teneufm
\scriptfont\eufmfam=\seveneufm \scriptscriptfont\eufmfam=\fiveeufm
\textfont\eusmfam=\teneusm\scriptfont\eusmfam=\seveneusm
\scriptscriptfont\eusmfam=\fiveeusm

\def\frak{\ifmmode\let\next\frak@\else
\def\next{\errmessage{Use\string\frak\space only in math
mode}}\fi\next}\def\frak@#1{{\frak@@{#1}}}
\def\frak@@#1{\fam\eufmfam#1}
\def\sh{\ifmmode\let\next\sh@\else
\def\next{\errmessage{Use\string\sh\space only in math
mode}}\fi\next}\def\sh@#1{{\sh@@{#1}}}
\def\sh@@#1{\fam\eusmfam#1}

\ifcase\@ptsize\font\tenmsa=msam10\font\sevenmsa=msam7
\font\fivemsa=msam5\font\tenmsb=msbm10
\font\sevenmsb=msbm7\font\fivemsb=msbm5\or \font\tenmsa=msam10
scaled\magstephalf \font\sevenmsa=msam7\font\fivemsa=msam5
\font\tenmsb=msbm10 scaled\magstephalf
\font\sevenmsb=msbm7\font\fivemsb=msbm5\or \font\tenmsa=msam10
scaled\magstep1\font\sevenmsa=msam7
\font\fivemsa=msam5\font\tenmsb=msbm10 scaled\magstep1
\font\sevenmsb=msbm7\font\fivemsb=msbm5\fi

\newfam\msafam\newfam\msbfam\textfont\msafam=\tenmsa
\scriptfont\msafam=\sevenmsa
\scriptscriptfont\msafam=\fivemsa\textfont\msbfam=\tenmsb
\scriptfont\msbfam=\sevenmsb \scriptscriptfont\msbfam=\fivemsb

\def\Bbb{\ifmmode\let\next\Bbb@\else
\def\next{\errmessage{Use\string\Bbb\space only in math
mode}}\fi\next}\def\Bbb@#1{{\Bbb@@{#1}}}
\def\Bbb@@#1{\fam\msbfam#1}\def\hexnumber@#1{\ifnum#1<10
\number#1\else\ifnum#1=10 A\else\ifnum#1=11 B\else\ifnum#1=12
C\else\ifnum#1=13 D\else\ifnum#1=14 E\else\ifnum#1=15
F\fi\fi\fi\fi\fi\fi\fi}
\def\msa@{\hexnumber@\msafam}\def\msb@{\hexnumber@\msbfam}
\mathchardef\square="0\msa@03

\makeatother
\newcommand\RR{{\Bbb R}}
\newcommand{\CC}{{\Bbb C}}

\begin{document}

\def\title#1{\centerline{\huge{#1}}}
\def\author#1{\centerline{\large{#1}}}
\def\address#1{\centerline{\it #1}}
\def\ack{{\bf Acknowledgments}$\quad$}
\def\Bibliography{\begin{thebibliography}}
\def\endbib{\end{thebibliography}}


\begin{titlepage}


\begin{center}

{\large\bf The Equivalence Postulate of Quantum Mechanics:\\ Main Theorems
\footnote{Contribution to the book on Quantum Trajectories, 
Edited by Pratim Chattaraj, Taylor\&Francis/CRC press.}}

\vspace{.999cm}

{\large Alon E. Faraggi$^1$ $\,$and$\,$ Marco Matone$^2$\\}

\vspace{.2in} {\it $^1$ Department of Mathematical Sciences,
University of Liverpool,
Liverpool L69 7ZL, UK\\
e-mail: faraggi@amtp.liv.ac.uk\\} \vspace{.02in} {\it $^2$
Department of Physics ``G. Galilei'' -- Istituto
Nazionale di Fisica Nucleare\\
University of Padova, Via Marzolo, 8 -- 35131 Padova, Italy\\
e-mail: matone@pd.infn.it\\}

\end{center}

\vspace{.333cm}

\begin{abstract} \noindent
We consider the two main theorems in the derivation of the Quantum
Hamilton--Jacobi Equation from the Equivalence Postulate (EP) of
quantum mechanics. The first one concerns a basic cocycle condition,
which holds in any dimension with Euclidean or Minkowski metrics
and implies a global conformal symmetry underlying the Quantum Hamilton--Jacobi Equation.
In one dimension such a condition fixes the Schwarzian
equation. The second theorem concerns
energy quantization which follows rigorously from consistency of the EP.

\end{abstract}

\noindent

\end{titlepage}\newpage
\setcounter{footnote}{0}
\renewcommand{\thefootnote}{\arabic{footnote}}

\section{HJ Equation and Coordinate Transformations}

The Hamilton--Jacobi (HJ) equation for a one
dimensional system is obtained by considering the canonical
transformation $(q,p)\to (Q,P)$ so that the old Hamiltonian $H$ maps to a trivialized one, that is
$\tilde H=0$. The old and new momenta are expressed in terms of
the generating function of such a transformation, the Hamilton's
principal function $p={\partial{\cal S}^{cl}\over\partial q}$, $P=cnst=-{\partial{\cal S}^{cl}\over
\partial Q}{|_{Q=cnst}}$
that satisfies the classical HJ equation $$
H\left(q,p={\partial{\cal S}^{cl}\over\partial q},t\right)+
{\partial{\cal S}^{cl}\over\partial t}=0\ . $$ In the case of a time
independent potential the time dependence in the Hamilton's
principal function ${\cal S}^{cl}$ is linear, that is
${\cal S}^{cl}(q,Q,t)={\cal S}_0^{cl}(q,Q)-Et$, with $E$ the energy of the
stationary state. It follows that ${\cal S}_0^{cl}$, called
Hamilton's characteristic function, or reduced action, satisfies the
Classical Stationary HJ Equation (CSHJE) $$
H\left(q,p={\partial{\cal S}_0^{cl}\over\partial q}\right)-E=0\ , $$
that is (${\cal W}(q)\equiv V(q)-E$) $$
{1\over2m}\left({\partial{\cal S}_0^{cl}\over\partial
q}\right)^2+{\cal W}=0\ .
$$

Note that the canonical transformation $(q,p)\to (Q,P)$ treats $p$ and $q$ as independent variables.
Following \cite{1} we now formulate a similar question to that
leading to the CSHJE, but considering the transformation on $q$, with the one on $p$ induced by the relation
$$ p={\partial{\cal
S}_0^{cl}\over\partial q}\ .
$$
More precisely, given a one dimensional system, with
time--independent potential (the higher dimensional time--dependent
case is considered in \cite{BFM}) we look for the coordinate
transformation $q\to q_0$ such that \be {\cal
S}_0^{cl}(q)\qquad\stackrel{Coord.\;Transf.}{\longleftrightarrow}\qquad\tilde
{\cal S}_0^{cl\,0}(q_0)\ , \label{staccorello2}\ee with $\tilde{\cal
S}_0^{ cl\,0}(q_0)$ denoting the reduced action of the system with
vanishing Hamiltonian. Note that in (\ref{staccorello2}) we required
that this transformation be an invertible one. This is an important
point since by compositions of the maps it follows that if for each
system there is a coordinate transformation leading to the trivial
state, then even two arbitrary systems are equivalent under
coordinate transformations. Imposing this apparently harmless
analogy immediately leads to rather peculiar properties of Classical
Mechanics (CM). First, it is clear that such an equivalence
principle cannot be satisfied in CM, in other words given two
arbitrary systems $a$ and $b$, the condition \be{\cal
S}_0^{cl\,b}(q_b)={\cal S}_0^{cl\,a}(q_a)\ , \label{staccc}\ee
cannot be generally satisfied. In particular, since
$$\tilde{\cal S}_0^{cl\,0}(q_0)=cnst\ ,$$ it is clear that
(\ref{staccorello2}) is a degenerate transformation. However, in
principle, by itself the failure of (\ref{staccc}) for arbitrary
systems would be a possible natural property. Nevertheless, a more
careful analysis shows that such a failure is strictly dependent on
the choice of the reference frame. This is immediately seen by
considering two free particles of mass $m_a$ and $m_b$ moving with
relative velocity $v$. For an observer at rest with respect to the
particle $a$ the two reduced actions are
$$ {\cal S}_0^{cl\,a}(q_a)=cnst\ ,\qquad{\cal S}_0^{cl\,b}(q_b)=m_bvq_b\ . $$ It
is clear that there is no way to have an equivalence under
coordinate transformations by setting ${\cal S}_0^{cl\,b}(q_b)={\cal
S}_0^{cl\,a}(q_a)$. This means that at the level of the reduced
action there is no coordinate transformation making the two systems
equivalent. However, note that this coordinate transformation exists
if we consider the same problem described by an observer in a frame
in which both particles have a non--vanishing velocity so that the
two particles are described by non--constant reduced actions.
Therefore, in CM, it is possible to connect different systems by a
coordinate transformation except in the case in which one of the
systems is described by a constant reduced action. This means that
in CM equivalence under coordinate transformations is frame
dependent. In particular, in the CSHJE description there is a
distinguished frame. This seems peculiar as on general grounds what
is equivalent under coordinate transformations in all frames should
remain so even in the one at rest.

\section{The Equivalence Postulate}

The above investigation already suggests that the concept of point
particle itself cannot be consistent with the equivalence under
coordinate transformations. In particular, it suggests that the
system where a particle is at rest does not exist at all. If this
would be the case, then the above critical situation would not occur
simply because the reduced action is never a constant. This should
reflect in two main features. First the classical concept of point
particle should be reconsidered, secondly the CSHJE should be
modified accordingly. A natural suggestion would be to consider
particles as a kind of string with a lower bound on the vibrating
modes in such a way that there is no way to define a system where
the particle is at rest. It should be observed that this kind of
string may differ from the standard one, rather its nature may be
related to the fact that in general relativity is impossible to
define the concept of relative stability of a system of particles.

In \cite{2} it was
suggested that quantum mechanics and gravity are intimately related. In particular, it was argued that
the quantum Hamilton-Jacobi equation of two free particles, which is attractive, may generate the gravitational potential.
This is a consequence of the fact that the quantum potential is always non-trivial even in the case of the free particle. It plays the role of intrinsic energy and may in fact be at the origin of fundamental interactions.

The unification of quantum mechanics and general relativity
is the central question of theoretical physics.
This problem hinges on the viability of the prevailing theories of
matter and interactions at the micro--scale, and of the
cosmos at the macro scale.
The Galilean paradigm of modern science drives
the search for a mathematical formulation of the quantum
gravity synthesis.
In such a context string theory provides an attempt for
a self--consistent mathematical  formulation of quantum
gravity.

String theory provides a perturbatively finite
$S$--matrix approach to the calculation of string
scattering amplitudes. Due to its
unique world--sheet properties, string theory
admits a discrete particle spectrum. It accommodates
the gauge bosons and fermion matter states that form the bedrock
of modern particle physics, as well as a massless spin 2 symmetric
state, which is interpreted as the gravitational force mediation field.
Consequently, string theory enables the construction of models
that admit the structures of the Standard Particle Model and
enable the development of a phenomenological approach to
quantum gravity. The state of the art in this regard is the
construction of Minimal Heterotic String Standard Models,
which produces in the observable Standard Model charged
sector solely the spectrum of the Minimal supersymmetric
Standard Model \cite{mshsm}.
Progress in the understanding of string theory
was obtained by
the observation that the five ten dimensional string
theories, as well as eleven dimensional supergravity,
can be connected by perturbative and non--perturbative
duality tranformations. However, this observation does not
provide a rigorous formulation of quantum gravity,
akin to the formulations of general relativity and quantum mechanics,
which follow from
the equivalence principle in the former and the probability
interpretation of the wave function in the later.

Let us start imposing the equivalence under coordinate
transformations. The key point is to consider, like in general
relativity, the (analogous of the) reduced action as a scalar field
under coordinate transformations.

We postulate that for any pair of one--particle states there exists
a field $\mathcal{S}_0$ such that \be
\mathcal{S}_0^b(q_b)=\mathcal{S}_0^a(q_a)\ , \label{abinitio}\ee is
well defined. We also require that, in a suitable limit,
$\mathcal{S}_0$ reduces to $\mathcal{S}_0^{cl}$. Eq.(\ref{abinitio})
can be considered as the scalar hypothesis. Since the conjugate
momentum is defined by $$ p_i={\partial\over\partial q^i}
\mathcal{S}_0(q)\ ,$$ it follows by (\ref{abinitio}) that the
conjugate momenta $p^a$ and $p^b$ are related by a coordinate
transformation \be p^b_i=\Lambda_i^{\,j}p^a_j\ , \label{dddff}\ee
where $\Lambda_i^{\,j}={\partial q_a^j/\partial q_b^i}$. Note that
we have the invariant \be p_i^bdq_b^i=p_i^adq_a^i\ .
\label{invva}\ee Since (\ref{abinitio}) holds for any pair of
one--particle states, we have ${\rm Det}\, \Lambda(q)\neq 0$,
$\forall q$.

The scalar hypothesis (\ref{abinitio}) implies that two
one--particle states are always connected by a coordinate
transformation, for such a reason we may equivalently consider
(\ref{abinitio}) as imposing an Equivalence Postulate (EP). In
particular, while in arbitrary dimension the coordinate
transformation is given by imposing (\ref{dddff}), in the one
dimensional case the scalar hypothesis implies
$$q_b={{\cal S}_0^b}^{-1}\circ{\cal S}_0^a(q_a) \ . $$

We now consider the consequences of the EP (\ref{abinitio}). Let us
denote by ${\cal H}$ the space of all possible ${\cal W}\equiv V-E$.
We also call $v$--transformations the ones leading from a system to
another. Eq.(\ref{abinitio}) is equivalent to require that

\vspace{.333cm}

\noindent {\it For each pair ${\cal W}^a,{\cal W}^b\in{\cal H}$,
there is a $v$--transformation such that} \be {\cal
W}^a(q)\longrightarrow{{\cal W}^a}^v(q^v)={\cal W}^b(q^v)\ .
\label{equivalence}\ee This implies that there always exists the
trivializing coordinate $q_0$ for which ${\cal
W}(q)\longrightarrow{\cal W}^0(q_0)$, where $$ {\cal W}^0(q_0)\equiv
0\ .
$$ In particular, since the inverse transformation should exist as
well, it is clear that the trivializing transformation should be
locally invertible. We will also see that since classically ${\cal
W}^0$ is a fixed point, implementation of (\ref{equivalence})
requires that ${\cal W}$ states transform inhomogeneously.

The fact that the EP cannot be consistently implemented in CM is
true in any dimension. To show this let us consider the coordinate
transformation induced by the identification \be {\cal S}_0^{cl\,
v}(q^v)={\cal S}_0^{cl}(q)\, . \label{pcowcOII}\ee Then note that
the CSHJE \be {1\over2m}\sum_{k=1}^D({\partial_{q_k}{\cal
S}_0^{cl}(q)})^2+{\cal W}(q)=0\ , \label{012}\ee provides a
correspondence between ${\cal W}$ and ${\cal S}_0^{cl}$ that we can
use to fix, by consistency, the transformation properties of ${\cal
W}$ induced by that of ${\cal S}_0^{cl}$. In particular, since
${\cal S}_0^{cl\, v}(q^v)$ must satisfy the CSHJE \be
{1\over2m}\sum_{k=1}^D(\partial_{q^{k\, v}}{\cal
S}_0^{cl\,v}(q^v))^2+{\cal W}^v(q^v)=0\ , \label{zummoloa30}\ee by
(\ref{pcowcOII}) we have \be {\partial{\cal
S}_0^{cl\,v}(q^v)\over\partial q^{k\, v}}=\Lambda_k^{\,
i}{\partial{\cal S}_0^{cl}(q)\over\partial q^i}\ .
\label{insomma}\ee Let us set $(p^v|p)={p^t\Lambda^t\Lambda p/
p^tp}$. By (\ref{012})--(\ref{insomma}), we have ${\cal
W}(q)\longrightarrow{\cal W}^v(q^v)=(p^v|p){\cal W}(q)$, so that $$
{\cal W}^0(q_0)\longrightarrow{\cal W}^v(q^v)=(p^v|p^0){\cal
W}^0(q_0)=0\ . $$ Thus we have \cite{1}

\vspace{.333cm}

\noindent {\it ${\cal W}$ states transform as quadratic
differentials under classical $v$--maps. It follows that ${\cal
W}^0$ is a fixed point in ${\cal H}$. Equivalently, in CM the space
${\cal H}$ cannot be reduced to a point upon factorization by the
classical $v$--transformations. Hence, the EP (\ref{equivalence})
cannot be consistently implemented in CM. This can be seen as the
impossibility of implementing covariance of CM under the coordinate
transformation defined by (\ref{pcowcOII}).}

\vspace{.333cm}

It is therefore clear that in order to implement the EP we have to
deform the CSHJE. As we will see, this requirement will determine
the equation for ${\cal S}_0$.

In Ref.\cite{1} the function ${\cal T}_0(p)$, defined as the
Legendre transform of the reduced action, was introduced $$ {\cal
T}_0(p)=q^kp_k-{\cal S}_0(q),\qquad{\cal S}_0(q)=p_kq^k-{\cal
T}_0(p)\ . $$ While ${\cal S}_0(q)$ is the momentum generating
function, its Legendre dual ${\cal T}_0(p)$ is the coordinate
generating function $$ p_k={\partial{\cal S}_0\over\partial
q_k},\qquad q_k={\partial{\cal T}_0\over\partial p_k}\ . $$ Note
that adding a constant to ${\cal S}_0$ does not change the dynamics.
Then, the most general differential equation ${\cal S}_0$ should
satisfy has the structure \be {\cal F}(\nabla{\cal S}_0,\Delta{\cal
S}_0,\ldots)=0\ . \label{traslazione}\ee \noindent Let us write down
Eq.(\ref{traslazione}) in the general form $$
{1\over2m}\sum_{k=1}^D({\partial_{q^k}{\cal S}_0(q)})^2+{\cal
W}(q)+Q(q)=0\ .
$$ The transformation properties of ${\cal W}+Q$ under the $v$--maps are
determined by the transformed equation \be
{1\over2m}\sum_{k=1}^D(\partial_{q^{k\, v}}{\cal
S}_0^v(q^v))^2/2m+{\cal W}^v(q^v)+Q^v(q^v)=0\ ,
\label{labella998}\ee so that \be {\cal
W}^v(q^v)+Q^v(q^v)=(p^v|p)\left[{\cal W}(q)+Q(q)\right]\ .
\label{yyyxxaa10bbbb}\ee

A basic guidance in deriving the differential equation for ${\cal
S}_0$ is that in some limit it should reduce to the CSHJE. In
\cite{1}\cite{BFM}\cite{2} it was shown that the parameter which
selects the classical phase is the Planck constant. Therefore, in
determining the structure of the $Q$ term we have to take into
account that in the classical limit \be \lim_{\hbar\to0}Q=0\ .
\label{classicoqezero}\ee

The only possibility to reach any other state ${\cal W}^v\ne0$
starting from ${\cal W}^0$ is that it transforms with an
inhomogeneous term. Namely as ${\cal W}^0\longrightarrow {\cal
W}^v(q^v)\ne0$, it follows that for an arbitrary ${\cal W}^a$ state
\be {\cal W}^v(q^v)=(p^v|p^a){\cal W}^a(q_a)+(q_a;q^v)\ ,
\label{azzoyyyxxaa10bbbb}\ee and by (\ref{yyyxxaa10bbbb}) \be
Q^v(q^v)=(p^v|p^a)Q^a(q_a)-(q_a;q^v)\ .
\label{azzo2yyyxxaa10bbbb}\ee Let us stress that the purely quantum
origin of the inhomogeneous term $(q_a;q^v)$ is particularly
transparent once one consider the compatibility between the
classical limit (\ref{classicoqezero}) and the transformation
properties of $Q$ in Eq.(\ref{azzo2yyyxxaa10bbbb}).

The ${\cal W}^0$ state plays a special role. Actually, setting
${\cal W}^a={\cal W}^0$ in Eq.(\ref{azzoyyyxxaa10bbbb}) yields $$
{\cal W}^v(q^v)=(q_0;q^v)\ , $$ so that, according to the EP
(\ref{equivalence}), all the states correspond to the inhomogeneous
part in the transformation of the ${\cal W}^0$ state induced by some
$v$--map.

Let us denote by $a,b,c,\ldots$ different $v$--transformations.
Comparing \be {\cal W}^b(q_b)=(p^b|p^a){\cal
W}^a(q_a)+(q_a;q_b)=(q_0;q_b)\ , \label{ganzate}\ee with the same
formula with $q_a$ and $q_b$ interchanged we have \be
(q_b;q_a)=-(p^a|p^b)(q_a;q_b)\ , \label{inparticolare}\ee in
particular $(q;q)=0$ More generally, imposing the commutative
diagram of maps $$
\begin{array}{c} {} \\ {} \\ A\end{array}
\begin{array}{c} {} \\ \nearrow \\ {} \end{array}
\begin{array}{c} B \\ {} \\ \longrightarrow \end{array}
\begin{array}{c} {} \\ \searrow \\ {} \end{array}
\begin{array}{c} {} \\ {} \\ C\end{array}
$$ that is comparing $$ {\cal W}^b(q_b)=(p^b|p^c){\cal W}^c(q_c)+(q_c;q_b)=
(p^b|p^a){\cal W}^a(q_a)+ (p^b|p^c)(q_a;q_c)+(q_c;q_b)\ , $$ with
(\ref{ganzate}), we obtain the basic cocycle condition \be
(q_a;q_c)=(p^c|p^b)\left[(q_a;q_b)+(q_b;q_c)\right]\ ,
\label{cociclo3}\ee which expresses the essence of the EP. In the
one dimensional case we have \be
(q_a;q_c)=\left(\partial_{q_c}q_b\right)^2(q_a;q_b)+(q_b;q_c)\ .
\label{inhomtrans}\ee It is well--known that this is satisfied by
the Schwarzian derivative. However, it turns out that it is
essentially the unique solution. More precisely \cite{1},

\vspace{.333cm}

\noindent {\bf Theorem 1.} {\it Eq.}(\ref{inhomtrans}) {\it defines
the Schwarzian derivative up to a multiplicative constant and a
coboundary term.}

\vspace{.333cm}

\noindent Since the differential equation for $\mathcal{S}_0$ should
depend only on $\partial_q^k\mathcal{S}_0$, $k\geq1$, it follows
that the coboundary term must be zero, so that \cite{1} $$
(q_a;q_b)=-{\beta^2\over4m}\{q_a,q_b\}\ , $$ where
$\{f(q),q\}=f'''/f'-3(f''/f')^2/2$ is the Schwarzian derivative and
$\beta$ is a nonvanishing constant that we identify with $\hbar$. As
a consequence, $\mathcal{S}_0$ satisfies the Quantum Stationary
Hamilton--Jacobi Equation (QSHJE) \cite{1} \be {1\over
2m}\left({\partial\mathcal{S}_0(q)\over\partial q}\right)^2+V(q)-E
+{\hbar^2\over4m}\{\mathcal{S}_0,q\}=0\ . \label{1Q}\ee Note that
$\psi={\mathcal{S}_0'}^{-1/2}\left(A
e^{-{i\over\hbar}\mathcal{S}_0}+Be^{{i\over\hbar}\mathcal{S}_0}\right)$
solves the Schr\"odinger Equation (SE) \be
\left(-{\hbar^2\over2m}{\partial^2\over\partial
q^2}+V\right)\psi=E\psi\ .\label{yz1xxxx4}\ee The ratio
$w=\psi^D/\psi$, where $\psi^D$ and $\psi$ are two real linearly
independent solutions of (\ref{yz1xxxx4}) is, in deep analogy with
uniformization theory, the {\it trivializing map} transforming any
$\mathcal{W}$ to $\mathcal{W}^0\equiv0$
\cite{1}\cite{Matone:1993tj}. This formulation
extends to higher dimension and to the
relativistic case as well \cite{1}\cite{BFM}.

Let $q_{-/+}$ be the lowest/highest $q$ for which $\mathcal{W}(q)$
changes sign, we have \cite{1}

\vspace{.333cm}

\noindent {\bf Theorem 2.} {\it If} \be
V(q)-E\geq\left\{\begin{array}{ll}P_-^2
>0\ ,&q<q_-\ ,\\ P_+^2 >0\ ,&q> q_+\ ,\end{array}\right.
\label{perintroasintoticopiumeno}\ee {\it then $w$ is a local
self--homeomorphism of $\hat{\RR}={\RR}\cup\{\infty\}$ if and only
if Eq.}(\ref{yz1xxxx4}) {\it has an $L^2(\RR)$ solution.}

\vspace{.333cm}

\noindent The crucial consequence is that since the QSHJE is defined
if and only if $w$ is a local self--homeomorphism of $\hat{\RR}$, it
follows that the QSHJE by itself implies energy quantization. We
stress that this result is obtained without any probabilistic
interpretation of the wave function.

\section{Proof of Theorem 1}

The main steps in proving theorem 1 are two lemmas \cite{1}. Let us
start observing that if the cocycle condition (\ref{inhomtrans}) is
satisfied by $(f(q);q)$, then this is still satisfied by adding a
coboundary term \be
(f(q);q)\longrightarrow(f(q);q)+(\partial_qf)^2G(f(q))-G(q)\ .
\label{coboundary}\ee Since $(Aq;q)$ evaluated at $q=0$ is
independent of $A$, we have \be 0=(q;q)=(q;q)_{|q=0}=(Aq;q)_{|q=0}\
. \label{valezero}\ee Therefore, if both $(f(q);q)$ and
(\ref{coboundary}) satisfy (\ref{inhomtrans}), then $G(0)=0$, which
is the unique condition that $G$ should satisfy. We now use
(\ref{traslazione}) to fix the ambiguity (\ref{coboundary}). First
of all observe that the differential equation we are looking for is
\be (q_0;q)={\cal W}(q)\ . \label{equazionewelook}\ee Then,
recalling that $q_0={\cal S}_0^{0^{\;-1}}\circ{\cal S}_0(q)$, we see
that a necessary condition to satisfy (\ref{traslazione}) is that
$(q_0;q)$ depends only on the first and higher derivatives of $q_0$.
This in turn implies that for any constant $B$ we have $(q_a+B;q_b)
=(q_a;q_b)$ that, together with (\ref{inparticolare}), gives \be
(q_a+B;q_b)=(q_a;q_b)=(q_a;q_b+B)\ . \label{1XuBtR}\ee Let $A$ be a
non--vanishing constant and set $h(A,q)=(Aq;q)$. By (\ref{1XuBtR})
we have $h(A,q+B)=h(A,q)$, that is $h(A,q)$ is independent of $q$.
On the other hand, by (\ref{valezero}) $h(A,0)=0$ that, together
with (\ref{inparticolare}), implies \be (Aq;q)=0=(q;Aq)\ .
\label{pt11}\ee Eq.(\ref{inhomtrans}) implies $(q_a;
Aq_b)=A^{-2}((q_a;q_b)-(Aq_b;q_b))$, so that by (\ref{pt11}) \be
(q_a;Aq_b)=A^{-2}(q_a;q_b)\ . \label{1Xw}\ee By
(\ref{inparticolare}) and (\ref{1Xw}) we have $$(Aq_a;q_b)=-A^{-2}(
\partial_{q_b}q_a)^2(q_b;Aq_a)
=-(\partial_{q_b}q_a)^2(q_b;q_a)= (q_a;q_b)\ ,$$ that is \be
(Aq_a;q_b)=(q_a;q_b)\ . \label{1Xu}\ee Setting
$f(q)=q^{-2}(q;q^{-1})$ and noticing that by (\ref{inparticolare})
and (\ref{1Xu}) $f(Aq)=-f(q^{-1})$, we obtain \be
(q;q^{-1})=0=(q^{-1};q)\ . \label{kenzoexenasarco}\ee Furthermore,
since by (\ref{inhomtrans}) and (\ref{kenzoexenasarco}) one has
$(q_a;q_b^{-1})=q_b^4(q_a;q_b)$, it follows that
$$(q_a^{-1};q_b)
=-\left(\partial_{q_b}q_a^{-1}\right)^2(q_b; q_a^{-1})
=-\left(\partial_{q_b}{q_a}\right)^2(q_b;q_a)=(q_a;q_b)\ ,
$$
so that \be (q_a^{-1};q_b)=(q_a;q_b)=q_b^{-4}(q_a;q_b^{-1})\ .
\label{pt22}\ee Since translations, dilatations and inversion are
the generators of the M\"obius group, it follows by
(\ref{1XuBtR})(\ref{1Xw})(\ref{1Xu}) and (\ref{pt22}) that

\vspace{.333cm}

\noindent {{\bf Lemma 1.}\it  Up to a coboundary term,
Eq.}(\ref{inhomtrans}) {\it implies} $$ (\gamma(q_a);q_b)=(q_a;q_b)\
, $$ $$
(q_a;\gamma(q_b))=\left(\partial_{q_b}\gamma(q_b)\right)^{-2}(q_a;q_b)\
, $$ where $\gamma(q)$ is an arbitrary $PSL(2,\CC)$ transformation.

\vspace{.333cm}

\noindent Now observe that since $(q_a;q_b)$ should depend only on
$\partial_{q_b}^kq_a$, $k\geq1$, we have \be (q+\epsilon
f(q);q)=c_1\epsilon f^{(k)}(q)+\mathcal{O}(\epsilon^2)\ ,
\label{preliminare1}\ee where $q_a=q+\epsilon f(q)$, $q\equiv q_b$
and $f^{(k)}\equiv\partial_q^kf$, $k\geq1$. Note that by lemma 1 and
(\ref{preliminare1}) $$(Aq+\epsilon Af(q);Aq)$$ \be =(q+\epsilon
f(q);Aq)=A^{-2}(q+\epsilon f(q);q)=A^{-2}c_1\epsilon
f^{(k)}(q)+\mathcal{O}(\epsilon^2)\ ,\label{preliminare2}\ee on the
other hand, setting $F(Aq)=Af(q)$, by (\ref{preliminare1})
$$(Aq+\epsilon Af(q);Aq)$$
$$=(Aq+\epsilon F(Aq);Aq)
=c_1\epsilon\partial_{Aq}^kF(Aq)+\mathcal{O}(\epsilon^2)=
A^{1-k}c_1\epsilon f^{(k)}(q)+\mathcal{O}(\epsilon^2)\ ,$$ that
compared with (\ref{preliminare2}) gives $k=3$. The above scaling
property generalizes to higher order contributions in $\epsilon$. In
particular, at order $\epsilon^n$ the quantity $(Aq+\epsilon
Af(q);Aq)$ is a sum of terms of the form
$$c_{i_1\ldots i_n}\partial_{Aq}^{i_1}\epsilon F(Aq)\cdot\cdot\cdot
\partial_{Aq}^{i_n}\epsilon F(Aq)
=c_{i_1\ldots i_n}\epsilon^nA^{n-\sum i_k}
f^{(i_1)}(q)\cdot\cdot\cdot f^{(i_n)}(q)\ ,$$ and by
(\ref{preliminare2}) $\sum_{k=1}^ni_k=n+2$. On the other hand, since
$(q_a;q_b)$ depends only on $\partial_{q_b}^kq_a$, $k\geq1$, we have
$$i_k\geq 1\ ,\qquad k\in[1,n]\ ,$$ so that either
$$i_k=3\ ,\qquad i_j=1\ ,\qquad
j\in[1,n]\ ,\qquad j\ne k\ ,$$ or $$i_k=i_j=2\ ,\qquad i_l=1\
,\qquad l\in[1,n]\ , \qquad l\ne k,\,l\ne j\ .$$ Hence
\be(q+\epsilon f(q);q)= \sum_{n=1}^\infty\epsilon^n\left(c_nf^{(3)}
f^{(1)^{n-1}}+d_nf^{(2)^2}f^{(1)^{n-2}}\right)\ ,\qquad d_1=0\ .
\label{preliminare4}\ee Let us now consider the transformations $$
q_b=v^{ba}(q_a)\ , \,\qquad q_c=v^{cb}(q_b)= v^{cb}\circ
v^{ba}(q_a)\ , \,\qquad q_c=v^{ca}(q_a)\ .$$ Note that
$v^{ab}=v^{ba^{-1}}$, and \be v^{ca}=v^{cb}\circ v^{ba}\ .
\label{vcacbba}\ee We can express these transformations in the form
$$ q_b=q_a+\epsilon^{ba}(q_a)\ ,$$ \be
q_c=q_b+\epsilon^{cb}(q_b)=q_b+
\epsilon^{cb}(q_a+\epsilon^{ba}(q_a))\ ,\label{epsilonabcvx}\ee
$$ q_c=q_a+\epsilon^{ca}(q_a)\ .$$  Since
$q_b=q_a-\epsilon^{ab}(q_b)$, we have $q_b=q_a-\epsilon^{ab}(q_a+
\epsilon^{ba}(q_a))$ that compared with $q_b=q_a+\epsilon^{ba}(q_a)$
yields $$ \epsilon^{ba}+\epsilon^{ab}\circ({\bf 1}+\epsilon^{ba})=0\
,
$$ where ${\bf 1}$ denotes the identity map. More generally,
Eq.(\ref{epsilonabcvx}) gives$$ \epsilon^{ca}(q_a)=
\epsilon^{cb}(q_b)+\epsilon^{ba}(q_a)=
\epsilon^{cb}(q_b)-\epsilon^{ab}(q_b)\ , $$ so that we obtain
(\ref{vcacbba}) with $v^{yx}={\bf 1}+\epsilon^{yx}$ \be
\epsilon^{ca}=\epsilon^{cb}\circ({\bf
1}+\epsilon^{ba})+\epsilon^{ba}= ({\bf 1}+\epsilon^{cb})\circ({\bf
1}+\epsilon^{ba})-{\bf 1}\ . \label{unpoininoFBT2}\ee Let us
consider the case in which $\epsilon^{yx}(q_x)=\epsilon f_{yx}
(q_x)$, with $\epsilon$ infinitesimal. At first--order in $\epsilon$
Eq.(\ref{unpoininoFBT2}) reads \be
\epsilon^{ca}=\epsilon^{cb}+\epsilon^{ba}\ ,
\label{unpoininoFBT3}\ee in particular,
$\epsilon^{ab}=-\epsilon^{ba}$. Since $(q_a;q_b)=c_1
{\epsilon^{ab}}'''(q_b)+{\cal O}^{ab}(\epsilon^2)$, where $'$
denotes the derivative with respect to the argument, we can use the
cocycle condition (\ref{inhomtrans}) to get $$
c_1{\epsilon^{ac}}'''(q_c)+{\cal O}^{ac}(\epsilon^2)$$ \be
=(1+{\epsilon^{bc}}'(q_c))^2\left(c_1{\epsilon^{ab}}'''(q_b)+{\cal
O}^{ab}(\epsilon^2) -c_1{\epsilon^{cb}}'''(q_b)-{\cal
O}^{cb}(\epsilon^2)\right)\ , \label{unpoininoFBT4}\ee that at
first--order in $\epsilon$ corresponds to (\ref{unpoininoFBT3}). We
see that $c_1\ne0$. For, if $c_1=0$, then by (\ref{unpoininoFBT4}),
at second--order in $\epsilon$ one would have \be {\cal
O}^{ac}(\epsilon^2)={\cal O}^{ab}(\epsilon^2)-{\cal
O}^{cb}(\epsilon^2)\ , \label{unpoininoFBT5}\ee which contradicts
(\ref{unpoininoFBT3}). In fact, by (\ref{preliminare4}) we have $$
{\cal
O}^{ab}(\epsilon^2)=c_2{\epsilon^{ab}}'''(q_b){\epsilon^{ab}}'(q_b)+
d_2{{\epsilon^{ab}}''}^2(q_b)+{\cal O}^{ab}(\epsilon^3)\ ,
$$ that together with (\ref{unpoininoFBT5})
provides a relation which cannot be consistent with
$\epsilon^{ac}(q_c)=\epsilon^{ab}(q_b)-\epsilon^{cb} (q_b)$. A
possibility is that $(q_a;q_b)=0$. However, this is ruled out by the
EP, so that $$ c_1\ne0\ .$$ Higher--order contributions due to a
non--vanishing $c_1$ are obtained by using
$$q_c=q_b+\epsilon^{cb}(q_b)\ , \qquad\quad \epsilon^{ac}
(q_c)=\epsilon^{ab}(q_b)-\epsilon^{cb}(q_b)\ ,$$ and
$\epsilon^{bc}(q_c)=- \epsilon^{cb}(q_b)$ in
$c_1\partial_{q_c}^3\epsilon^{ac}(q_c)$ and in
$$c_1\left(2\partial_{q_c}\epsilon^{bc}(q_c)+{\partial_{q_c}\epsilon^{bc}
(q_c)}^2\right)\partial^3_{q_b}\left({\epsilon^{ab}}(q_b)-{\epsilon^{cb}}
(q_b)\right)\ .$$ Note that one can also consider the case in which
both the first-- and second--order contributions to $(q_a;q_b)$ are
vanishing. However, this possibility is ruled out by a similar
analysis. In general, one has that if the first non--vanishing
contribution to $(q_a;q_b)$ is of order $\epsilon^n$, $n\geq2$,
then, unless $(q_a;q_b)=0$, the cocycle condition (\ref{inhomtrans})
cannot be consistent with the linearity of (\ref{unpoininoFBT3}).
Observe that we proved that $c_1\ne0$ is a necessary condition for
the existence of solutions $(q_a;q_b)$ of the cocycle condition
(\ref{inhomtrans}), depending only on the first and higher
derivatives of $q_a$. Existence of solutions follows from the fact
that the Schwarzian derivative $\{q_a,q_b\}$ solves
(\ref{inhomtrans}) and depends only on the first and higher
derivatives of $q_a$.

The fact that $c_1=0$ implies $(q_a;q_b)=0$, can be also seen by
explicitly evaluating the coefficients $c_n$ and $d_n$. These can be
obtained using the same procedure considered above to prove that
$c_1\ne0$. Namely, inserting the expansion (\ref{preliminare4}) in
(\ref{inhomtrans}) and using $q_c=q_b+ \epsilon^{cb}(q_b)$,
$\epsilon^{ac}(q_c)=\epsilon^{ab}(q_b)-\epsilon^{cb} (q_b)$ and
$\epsilon^{bc}(q_c)=-\epsilon^{cb}(q_b)$, we obtain \be
c_n=(-1)^{n-1}c_1\ ,\qquad d_n={3\over2}(-1)^{n-1}(n-1)c_1\ ,
\label{cenneedennexf}\ee which in fact are the coefficients one
obtains expanding $c_1\{q+\epsilon f(q),q\}$. However, we now use
only the fact that $c_1\ne0$, as the relation $(q+\epsilon
f(q);q)=c_1\{q+\epsilon f(q),q\}$ can be proved without making the
calculations leading to (\ref{cenneedennexf}). Summarizing, we have

\vspace{.333cm}

\noindent {{\bf Lemma 2.}\it  If $$q_a=q_b+\epsilon^{ab}(q_b)\ ,$$
the unique solution of Eq.}(\ref{inhomtrans}){\it, depending only on
the first and higher derivatives of $q_a$, is}
$$
(q_a;q_b)=c_1{\epsilon^{ab}}'''(q_b)+\mathcal{O}^{ab}(\epsilon^2)\
,\qquad c_1\ne0\ . $$

\noindent

\vspace{.333cm}

It is now easy to prove that, up to a multiplicative constant and a
coboundary term, the Schwarzian derivative is the unique solution of
the cocycle condition (\ref{inhomtrans}). Let us first note that
$$
[q_a;q_b]=(q_a;q_b)-c_1\{q_a;q_b\}\ ,
$$
satisfies the cocycle condition
$$
[q_a;q_c]=\left({\partial_{q_c}q_b}\right)^2\left([q_a;q_b]-[q_c;q_b]
\right)\ .
$$
In particular, since both $(q_a;q_b)$ and $\{q_a;q_b\}$ depend only
on the first and higher derivatives of $q_a$, we have, as in the
case of $(q+\epsilon f(q);q)$, that
$$
[q+\epsilon f(q);q]=\tilde c_1\epsilon f^{(3)}(q)+{\cal
O}(\epsilon^2)\ ,
$$
where either $\tilde c_1\ne0$ or $[q+\epsilon f(q);q]=0$. However,
since $\{q+\epsilon f(q);q\}=\epsilon f^{(3)}(q)+{\cal
O}(\epsilon^2)$ and $(q+\epsilon f(q);q)=\epsilon f^{(3)}(q)+{\cal
O}(\epsilon^2)$, we have $\tilde c_1=0$ and the Lemma yields
$[q+\epsilon f(q);q]=0$. Therefore, we have that the EP univocally
implies that
$$
(q_a;q_b)=-{\beta^2\over4m}\{q_a,q_b\}\ ,
$$
where for convenience we replaced $c_1$ by $-\beta^2/4m$. This
concludes the proof of theorem 1.

We observe that despite some claims \cite{Ovsienko1}, we have not be
able to find in the literature a complete and close proof of the
above theorem (see also \cite{DMS}). We thank D.B. Fuchs for a
bibliographic comment concerning the above theorem.

In deriving the equivalence of states we considered the case of
one--particle states with identical masses. The generalization to
the case with different masses is straightforward. In particular,
the right hand side of Eq.(\ref{inhomtrans}) gets multiplied by
$m_b/m_a$, so that the cocycle condition becomes
$$
m_a(q_a;q_c)=m_a\left(\partial_{q_c}q_b\right)^2(q_a;q_b)+m_b(q_b;q_c)\
, $$ explicitly showing that the mass appears in the denominator and
that it refers to the label in the first entry of $(\cdot\,;\cdot)$,
that is \be (q_a;q_b)=-{\hbar^2\over 4m_a}\{q_a;q_b\}\ .
\label{bella}\ee The QSHJE (\ref{1Q}) follows almost immediately by
(\ref{bella}) \cite{1}.

The above investigation may be applied to CFT. Let us consider a
local conformal transformation of the stress tensor in a 2D CFT. The
infinitesimal variation of $T$ is given by \be \delta_\epsilon
T(w)=-{1\over12}c\partial_w^3\epsilon(w)-2T(w)\partial_w\epsilon(w)-\epsilon(w)\partial_wT(w)
\ , \label{hJ1}\ee where $c$ is the central charge. The finite
version of such a transformation is \be \tilde T(w)=(\partial_w
z)^2T(z)+{c\over12}\{w,z\} \ . \label{hJ2}\ee While it is immediate
to see that (\ref{hJ2}) implies (\ref{hJ1}), the viceversa is not
evident. A possible way to prove (\ref{hJ2}) is just to set \be
\tilde T(w)=(\partial_w z)^2T(z)+k(w;z) \ , \label{hJ3}\ee and then
to impose the cocycle condition which will show that $(w;z)$ is
proportional to $\{w,z\}$.  Comparison with the infinitesimal
transformation (\ref{hJ1}) fixes the constant $k$.

In \cite{BFM} it has been shown that the cocycle condition fixes the
higher dimensional version of the Schwarzian derivative. In this
respect we observe that its definition seems an open question in
mathematical literature. While in the one dimensional case the QSHJE
reduces to a unique differential equation, this is not immediate in
the higher dimensional case. However, it turns out that such a
reduction exists upon introducing an antisymmetric tensor \cite{BFM}
(in this respect it is worth noticing that some author introduces a
connection to define the higher dimensional Schwarzian derivative).

A basic feature of the cocycle condition is that it implies, as it
should, the higher dimensional M\"obious invariance with respect to
$q_a$ in $(q_a;q_b)$ (with similar properties with respect to
$q_b$). In particular, in \cite{BFM} it has been shown that \be
(q^a;q^b)=-{\hbar^2\over2m}\left[(p^b|p^a) {\Delta^aR^a\over
R^a}-{\Delta^bR^b\over R^b}\right]\ . \label{cocicloide}\ee It would
be interesting to consider such a definition in the context of the
transformation properties of the stress tensor in higher dimensional
CFTs.

\section{Proof of Theorem 2}

The QSHJE is equivalent to \be \{w,q\}=-{4m\over\hbar^2}{\cal W}(q)\
, \label{cosicchesothat}\ee where $w=\psi^D/\psi$ with $\psi^D$ and
$\psi$ two real linearly independent solutions of the Schr\"odinger
equation. Existence of this equation requires some conditions on the
continuity properties of $w$ and its derivatives. Since the QSHJE is
the consequence of the EP, we can say that the EP imposes some
constraints on $w=\psi^D/\psi$. These constraints are nothing but
the existence of the QSHJE (\ref{1Q}) or, equivalently, of
Eq.(\ref{cosicchesothat}). That is, implementation of the EP imposes
that $\{w,q\}$ exists, so that \be w\ne cnst,\;w\in C^2(\RR)
\;and\;\partial_q^2w\; differentiable \;on\;\RR\ .
\label{ccnndraft}\ee These conditions are not complete. The reason
is that, as we have seen, the implementation of the EP requires that
the properties of the Schwarzian derivative be satisfied. Actually,
its very properties, derived from the EP, led to the identification
$(q_a;q_b)=-\hbar^2 \{q_a,q_b\}/4m$. Therefore, in order to
implement the EP, the transformation properties of the Schwarzian
derivative and its symmetries must be satisfied. In deriving the
transformation properties of $(q_a;q_b)$ we noticed how, besides
dilatations and translations, there is a highly non--trivial
symmetry such as that under inversion. Therefore, we have that
(\ref{cosicchesothat}) must be equivalent to
$$
\{w^{-1},q\}=-{4m\over\hbar^2}{\cal W}(q)\ .
$$
A property of the Schwarzian derivative is duality between its
entries \be \{w,q\}=-\left({\partial w\over\partial
q}\right)^2\{q,w\}\ . \label{pksxjq}\ee This shows that the
invariance under inversion of $w$ reflects in the invariance, up to
a Jacobian factor, under inversion of $q$. That is
$\{w,q^{-1}\}=q^4\{w,q\}$, so that the QSHJE (\ref{cosicchesothat})
can be written in the equivalent form \be
\{w,q^{-1}\}=-{4m\over\hbar^2}q^4{\cal W}(q)\ . \label{oidjwI939}\ee
In other words, starting from the EP one can arrive to either
Eq.(\ref{cosicchesothat}) or Eq.(\ref{oidjwI939}). The consequence
of this fact is that since under
$$
q\rightarrow {1\over q}\ ,
$$
$0^{\pm}$ maps to $\pm\infty$, we have to extend (\ref{ccnndraft})
to the point at infinity. In other words, (\ref{ccnndraft}) should
hold on the extended real line $\hat\RR =\RR\cup\{\infty\}$. This
aspect is related to the fact that the M\"obius transformations,
under which the Schwarzian derivative transforms as a quadratic
differential, map circles to circles. We stress that we are
considering the systems defined on $\RR$ and not $\hat\RR$. What
happens is that the existence of the QSHJE forces us to impose
smoothly joining conditions even at $\pm\infty$, that is
(\ref{ccnndraft}) must be extended to \be w\ne cnst,\;w\in
C^2(\hat\RR)\;and\;\partial_q^2w\;differentiable\;on\;\hat\RR\ .
\label{ccnn}\ee One may easily check that $w$ is a M\"obius
transformation of the trivializing map \cite{1}. Therefore,
Eq.(\ref{pksxjq}), which is defined if and only if $w(q)$ can be
inverted, that is if $\partial_q w\ne 0$, $\forall q\in\RR$, is a
consequence of the cocycle condition (\ref{cociclo3}). By
(\ref{oidjwI939}) we see that also local univalence should be
extended to $\hat\RR$. This implies the following joining condition
at spatial infinity \be w(-\infty)=\left\{\begin{array}{ll}
w(+\infty)\ , & {\rm for}\quad w(-\infty)\ne\pm \infty\ ,\\
-w(+\infty)\ ,& {\rm for}\quad w(-\infty)=\pm\infty\
.\end{array}\right. \label{specificandoccnn}\ee As illustrated by
the non--univalent function $w=q^2$, the apparently natural choice
$w(-\infty)=w(+\infty)$, one would consider also in the $w(-\infty)=
\pm\infty$ case, does not satisfy local univalence.

We saw that the EP implied the QSHJE (\ref{1Q}). However, although
this equation implies the SE, we saw that there are aspects
concerning the canonical variables which arise in considering the
QSHJE rather than the SE. In this respect a natural question is
whether the basic facts of QM also arise in our formulation. A basic
point concerns a property of many physical systems such as energy
quantization. This is a matter of fact beyond any interpretational
aspect of QM. Then, as we used the EP to get the QSHJE, it is
important to understand how energy quantization arises in our
approach. According to the EP, the QSHJE contains all the possible
information on a given system. Then, the QSHJE itself should be
sufficient to recover the energy quantization including its
structure. In the usual approach the quantization of the spectrum
arises from the basic condition that in the case in which
$\lim_{q\to\pm\infty}{\cal W}>0$, the wave--function should vanish
at infinity. Once the possible solutions are selected, one also
imposes the continuity conditions whose role in determining the
possible spectrum is particularly transparent in the case of
discontinuous potentials. For example, in the case of the potential
well, besides the restriction on the spectrum due to the $L^2(\RR)$
condition for the wave--function (a consequence of the probabilistic
interpretation of the wave--function), the spectrum is further
restricted by the smoothly joining conditions. Since the SE contains
the term $\partial_q^2\psi$, the continuity conditions correspond to
an existence condition for this equation. On the other hand, also in
this case, the physical reason underlying this request is the
interpretation of the wave--function in terms of probability
amplitude. Actually, strictly speaking, the continuity conditions
come from the continuity of the probability density $\rho=|\psi|^2$.
This density should also satisfy the continuity equation
$\partial_t\rho+\partial_qj=0$, where $j=i\hbar(\psi
\partial_q\bar\psi-\bar\psi\partial_q\psi)/2m$. Since for stationary states
$\partial_t\rho=0$, it follows that in this case $j=cnst$.
Therefore, in the usual formulation, it is just the interpretation
of the wave--function in terms of probability amplitude, with the
consequent meaning of $\rho$ and $j$, which provides the physical
motivation for imposing the continuity of the wave--function and of
its first derivative.

Now observe that in our formulation the continuity conditions arise
from the QSHJE. In fact, (\ref{ccnn}) implies continuity of
$\psi^D$, $\psi$, with $\partial_q\psi^D$ and $\partial_q\psi$
differentiable, that is \be
EP\;\rightarrow\;(\psi^D,\psi)\;continuous\; and
\;(\psi^{D'},\psi')\;differentiable\ .
\label{equivalenzaederivata}\ee

In the following we will see that if $V(q)>E$, $\forall q\in\RR$,
then there are no solutions such that the ratio of two real linearly
independent solutions of the SE corresponds to a local
self--homeomorphism of $\hat\RR$. The fact that this is an
unphysical situation can be also seen from the fact that the case
$V>E$, $\forall q\in\RR$, has no classical limit. Therefore, if
$V>E$ both at $-\infty$ and $+\infty$, a physical situation requires
that there are at least two points where $V-E=0$. More generally, if
the potential is not continuous, $V(q)-E$ should have at least two
turning points. Let us denote by $q_-$ ($q_+$) the lowest (highest)
turning point. Note that by (\ref{perintroasintoticopiumeno}) we
have
$$
\int^{-\infty}_{q_-}dx\kappa(x)=-\infty\
,\qquad\quad\int^{+\infty}_{q_+}dx\kappa(x)=+ \infty\ ,
$$
where $\kappa=\sqrt{2m(V-E)}/\hbar$. Before going further, let us
stress that what we actually need to prove is that, in the case
(\ref{perintroasintoticopiumeno}), the joining condition
(\ref{specificandoccnn}) requires that the corresponding SE has an
$L^2(\RR)$ solution. Observe that while (\ref{ccnn}), which however
follows from the EP, can be recognized as the standard condition
(\ref{equivalenzaederivata}), the other condition
(\ref{specificandoccnn}), which still follows from the existence of
the QSHJE, and therefore from the EP, is not directly recognized in
the standard formulation. Since this leads to energy quantization,
while in the usual approach one needs one more assumption, we see
that there is quite a fundamental difference between the QSHJE and
the SE. We stress that (\ref{ccnn}) and (\ref{specificandoccnn})
guarantee that $w$ is a local self--homeomorphism of $\hat\RR$.

Let us first show that the request that the corresponding SE has an
$L^2(\RR)$ solution is a sufficient condition for $w$ to satisfy
(\ref{specificandoccnn}). Let $\psi\in L^2(\RR)$ and denote by
$\psi^D$ a linearly independent solution. As we will see, the fact
that $\psi^D\not\propto\psi$ implies that if $\psi\in L^2(\RR)$,
then $\psi^D\notin L^2(\RR)$. In particular, $\psi^D$ is divergent
both at $q=-\infty$ and $q=+\infty$. Let us consider the real ratio
$$
w={A\psi^D+B\psi\over C\psi^D+D\psi}\ ,
$$
where $AD-BC\ne 0$. Since $\psi\in L^2(\RR)$, we have \be
\lim_{q\rightarrow\pm\infty}w=\lim_{q\rightarrow\pm\infty}
{A\psi^D+B\psi\over C\psi^D+D\psi}={A\over C}\ ,
\label{arbitygvxy}\ee that is $w(-\infty)=w(+\infty)$. In the case
in which $C=0$ we have
$$
\lim_{q\rightarrow\pm\infty}w=\lim_{q\rightarrow\pm\infty}{A\psi^D
\over D\psi}=\pm\epsilon\cdot\infty\ ,
$$
where $\epsilon=\pm1$. The fact that ${A\psi^D/D\psi}$ diverges for
$q\to\pm\infty$ follows from the mentioned properties of $\psi^D$
and $\psi$. It remains to check that if
$\lim_{q\to-\infty}{A\psi^D/D\psi}=-\infty$, then
$\lim_{q\to+\infty}{A\psi^D/D\psi}=+\infty$, and vice versa. This
can be seen by observing that
$$
\psi^D(q)=c\psi(q)\int^q_{q_0}dx\psi^{-2}(x)+d\psi(q)\ ,
$$
$c\in\RR\backslash\{0\}$, $d\in\RR$. Since $\psi\in L^2(\RR)$ we
have $\psi^{-1}\not\in L^2(\RR)$ and
$\int^{+\infty}_{q_0}dx\psi^{-2}=+\infty$,
$\int^{-\infty}_{q_0}dx\psi^{-2}=-\infty$, implying that
$\psi^D(-\infty)/\psi
(-\infty)=-\epsilon\cdot\infty=-\psi^D(+\infty)/\psi(+\infty)$,
where $\epsilon={\rm sgn}\,c$.

We now show that the existence of an $L^2(\RR)$ solution of the SE
is a necessary condition to satisfy the joining condition
(\ref{specificandoccnn}). We give two different proofs of this, one
is based on the WKB approximation while the other one uses Wronskian
arguments. In the WKB approximation, we have \be
\psi={A_-\over\sqrt{\kappa}}e^{-\int^q_{q_-}dx\kappa}
+{B_-\over\sqrt{\kappa}}e^{\int^q_{q_-}dx\kappa},\quad q\ll q_-\ ,
\label{Pantani1}\ee and \be
\psi={A_+\over\sqrt{\kappa}}e^{-\int^q_{q_+}dx\kappa}
+{B_+\over\sqrt{\kappa}}e^{\int^q_{q_+}dx\kappa},\quad q\gg q_+\ .
\label{Pantani2}\ee In the same approximation, a linearly
independent solution has the form
$$
\psi^D={A_-^D\over\sqrt{\kappa}}e^{-\int^q_{q_-}dx\kappa}
+{B_-^D\over {\kappa}}e^{\int^q_{q_-}dx\kappa},\quad q\ll q_-\ .
$$
Similarly, in the $q\gg q_+$ region we have
$$
\psi^D={A_+^D\over\sqrt{\kappa}}e^{-\int^q_{q_+}dx\kappa}
+{B_+^D\over\sqrt{\kappa}}e^{\int^q_{q_+}dx\kappa},\quad q\gg q_+\ .
$$
Note that (\ref{Pantani1}) and (\ref{Pantani2}) are derived by
solving the differential equations corresponding to the WKB
approximation for $q\ll q_-$ and $q\gg q_+$, so that the
coefficients of $\kappa^{-1/2}\exp\pm\int^q_{q_-}dx \kappa$, {\it
e.g.} $A_-$ and $B_-$ in (\ref{Pantani1}), cannot be simultaneously
vanishing. In particular, the fact that $\psi^D\not\propto\psi$
yields \be A_-B_-^D-A_-^DB_-\ne 0\ ,\qquad A_+B_+^D-A_+^DB_+\ne 0\ .
\label{PantaniGirodItaliaeTour}\ee Let us now consider the case in
which, for a given $E$ satisfying (\ref{perintroasintoticopiumeno}),
any solution of the corresponding SE diverges at least at one of the
two spatial infinities, that is \be \lim_{q\rightarrow +\infty}
(|\psi(-q)|+|\psi(q)|)=+\infty\ . \label{caruccioe}\ee This implies
that there is a solution diverging both at $q=-\infty$ and $q=+
\infty$. In fact, if two solutions $\psi_1$ and $\psi_2$ satisfy
$\psi_1(- \infty)=\pm\infty$, $\psi_1(+\infty)\ne\pm\infty$ and
$\psi_2(-\infty)\ne\pm \infty$, $\psi_2(+\infty)=\pm\infty$, then
$\psi_1+\psi_2$ diverges at $\pm \infty$. On the other hand,
(\ref{PantaniGirodItaliaeTour}) rules out the case in which all the
solutions in their WKB approximation are divergent only at one of
the two spatial infinities, say $-\infty$. Since, in the case
(\ref{perintroasintoticopiumeno}), a solution which diverges in the
WKB approximation is itself divergent (and vice versa), we have that
in the case (\ref{perintroasintoticopiumeno}), the fact that all the
solutions of the SE diverge only at one of the two spatial
infinities cannot occur.

Let us denote by $\psi$ a solution which is divergent both at
$-\infty$ and $+\infty$. In the WKB approximation this means that
both $A_-$ and $B_+$ are non--vanishing, so that
$$
\psi{}_{\;\stackrel{\sim}{q\rightarrow-\infty}\;}{A_-\over\sqrt\kappa}e^{-
\int^q_{q_-}dx\kappa},\quad\qquad
\psi{}_{\;\stackrel{\sim}{q\rightarrow+\infty}
\;}{B_+\over\sqrt{\kappa}}e^{\int^q_{q_+} dx\kappa}\ .
$$
The asymptotic behavior of the ratio $\psi^D/\psi$ is given by
$$
\lim_{q\rightarrow-\infty}{\psi^D\over\psi}={A_-^D\over A_-}\
,\qquad\quad \lim_{q\rightarrow+\infty}{\psi^D\over\psi}={B_+^D\over
B_+}\ .
$$
Note that since in the case at hand any divergent solution also
diverges in the WKB approximation, we have that (\ref{caruccioe})
rules out the case $A^D_-= B_+^D=0$. Let us then suppose that either
$A_-^D=0$ or $B_+^D=0$. If $A_-^D=0$, then $w(-\infty)=0\ne
w(+\infty)$. Similarly, if $B_+^D=0$, then $w(+\infty)=0 \ne
w(-\infty)$. Hence, in this case $w$, and therefore the trivializing
map, cannot satisfy (\ref{specificandoccnn}). On the other hand,
also in the case in which both $A_-^D$ and $B_+^D$ are
non--vanishing, $w$ cannot satisfy Eq.(\ref{specificandoccnn}). For,
if $A_-^D/A_-=B_+^D/B_+$, then
$$
\phi=\psi-{A_-\over A^D_-}\psi^D=\psi-{B_+\over B^D_+}\psi^D\ ,
$$
would be a solution of the SE whose WKB approximation has the form
$$
\phi={B_-\over\sqrt{\kappa}}e^{\int^q_{q_-}dx\kappa},\qquad q\ll
q_-\ ,
$$
and
$$
\phi={A_+\over\sqrt{\kappa}}e^{-\int^q_{q_+}dx\kappa},\qquad q\gg
q_+\ .
$$
Hence, if $A_-^D/A_-=B_+^D/B_+$, then there is a solution whose WKB
approximation vanishes both at $-\infty$ and $+\infty$. On the other
hand, we are considering the values of $E$ satisfying
Eq.(\ref{perintroasintoticopiumeno}) and for which any solution of
the SE has the property (\ref{caruccioe}). This implies that no
solutions can vanish both at $-\infty$ and $+\infty$ in the WKB
approximation. Hence
$$
{A_-^D\over A_-}\ne{B_+^D\over B_+}\ ,
$$
so that $w(-\infty)\ne w(+\infty)$. We also note that not even the
case $w(- \infty)=\pm\infty=-w(+\infty)$ can occur, as this would
imply that $A_-=B_+=0$, which in turn would imply, against the
hypothesis, that there are solutions vanishing at $q=\pm\infty$.
Hence, if for a given $E$ satisfying
(\ref{perintroasintoticopiumeno}), any solution of the corresponding
SE diverges at least at one of the two spatial infinities, we have
that the trivializing map has a discontinuity at $q=\pm \infty$. As
a consequence, the EP cannot be implemented in this case so that
this value $E$ cannot belong to the physical spectrum.

Therefore, the physical values of $E$ satisfying
(\ref{perintroasintoticopiumeno}) are those for which there are
solutions which are divergent neither at $-\infty$ nor at $+\infty$.
On the other hand, from the WKB approximation and
(\ref{perintroasintoticopiumeno}), it follows that the
non--divergent solutions must vanish both at $-\infty$ and
$+\infty$. It follows that the only energy levels satisfying the
property (\ref{perintroasintoticopiumeno}), which are compatible
with the EP, are those for which there exists the solution vanishing
both at $\pm\infty$. On the other hand, solutions vanishing as
$\kappa^{-1/2}\exp\int^q_{q_-}dx \kappa$ at $-\infty$ and
$\kappa^{-1/2}\exp-\int^q_{q_+}dx\kappa$ at $+\infty$, with
$P^2_\pm>0$, cannot contribute with an infinite value to
$\int^{+\infty}_{-\infty}dx\psi^2$. The reason is that existence of
the QSHJE requires that $\{e^{{2i\over\hbar}{\cal S}_0},q\}$ be
defined and this, in turn, implies that any solution of the SE must
be continuous. On the other hand, since $\psi$ is continuous, and
therefore finite also at finite values of $q$, we have
$\int^{q_b}_{q_a}dx\psi^2<+\infty$ for all finite $q_a$ and $q_b$.
In other words, the only possibility for a continuous function to
have a divergent value of $\int^{+ \infty}_{-\infty}dx\psi^2$ comes
from its behavior at $\pm\infty$. Therefore, since the
implementation of the EP in the case
(\ref{perintroasintoticopiumeno}) requires that the corresponding
$E$ should admit a solution with the behavior
$$
\psi{}_{\;\stackrel{\sim}{q\rightarrow-\infty}\;}{A_-\over\sqrt{\kappa}}
e^{\int^q_{q_-}dx\kappa},\qquad\psi{}_{\;\stackrel{\sim}{q\rightarrow+
\infty}\;}{B_+\over\sqrt{\kappa}}e^{-\int^q_{q_+}dx\kappa}\ ,
$$
we have the following basic fact

\vspace{.333cm}

\noindent {\it The values of $E$ satisfying \be
V(q)-E\geq\left\{\begin{array}{ll}P_-^2>0\ ,&q<q_-\ ,\\
P_+^2>0\ ,&q>q_+\ ,
\end{array}\right.
\label{asintoticopiumenofgt}\ee are physically admissible if and
only if the corresponding SE has an $L^2(\RR)$ solution. }

\vspace{.333cm}

We now give another proof of the fact that if ${\cal W}$ is of the
type (\ref{asintoticopiumenofgt}), then the corresponding SE must
have an $L^2(\RR)$ solution in order to satisfy
(\ref{specificandoccnn}). In particular, we will show that this is a
necessary condition. That this is sufficient has been already proved
above.

By Wronskian arguments, which can be found in Messiah's book
\cite{Messiah}, imply that if $V(q)-E\geq P_+^2>0$, $q>q_+$, then as
$q\rightarrow +\infty$, we have ($P_+>0$)

\begin{itemize}
\item[{\bf --}]{There is a solution of the SE that vanishes at least as
$e^{-P_+q}$.}
\item[{\bf --}]{Any other linearly independent solution diverges at least as
$e^{P_+q}$.}
\end{itemize}

\noindent Similarly, if $V(q)-E\geq P_-^2>0$, $q<q_-$, then as
$q\rightarrow- \infty$, we have ($P_->0$)

\begin{itemize}
\item[{\bf --}]{There is a solution of the SE that vanishes at least as
$e^{P_-q}$.}
\item[{\bf --}]{Any other linearly independent solution diverges at least as
$e^{-P_-q}$.}
\end{itemize}

\noindent These properties imply that if there is a solution of the
SE in $L^2(\RR)$, then any solution is either in $L^2(\RR)$ or
diverges both at $-\infty$ and $+\infty$. Let us show that the
possibility that a solution vanishes only at one of the two spatial
infinities is ruled out. Suppose that, besides the $L^2(\RR)$
solution, which we denote by $\psi_1$, there is a solution $\psi_2$
which is divergent only at $+\infty$. On the other hand, the above
properties show that there exists also a solution $\psi_3$ which is
divergent at $-\infty$. Since the number of linearly independent
solutions of the SE is two, we have $\psi_3=A \psi_1+B\psi_2$.
However, since $\psi_1$ vanishes both at $-\infty$ and $+\infty$, we
see that $\psi_3=A\psi_1+B\psi_2$ can be satisfied only if $\psi_2$
and $\psi_3$ are divergent both at $-\infty$ and $+\infty$. This
fact and the above properties imply that

\vspace{.333cm}

\noindent {\it If the SE has an $L^2(\RR)$ solution, then any
solution has two possible asymptotics}

\begin{itemize}
\item[{\bf --}]{Vanishes both at $-\infty$ and $+\infty$ at least as $e^{P_-q}$
and $e^{-P_+q}$ respectively.}
\item[{\bf --}]{Diverges both at $-\infty$ and $+\infty$ at least as $e^{-P_-q}$
and $e^{P_+q}$ respectively.}
\end{itemize}

\vspace{.333cm}

\noindent Similarly, we have

\vspace{.333cm}

\noindent {\it If the SE does not admit an $L^2(\RR)$ solution, then
any solution has three possible asymptotics}

\begin{itemize}
\item[{\bf --}]{Diverges both at $-\infty$ and $+\infty$ at least as $e^{-P_-q}$
and $e^{P_+q}$ respectively.}
\item[{\bf --}]{Diverges at $-\infty$ at least as $e^{-P_-q}$ and vanishes at
$+\infty$ at least as $e^{-P_+q}$.}
\item[{\bf --}]{Vanishes at $-\infty$ at least as $e^{P_-q}$ and diverges at
$+\infty$ at least as $e^{P_+q}$.}
\end{itemize}

\vspace{.333cm}

\noindent Let us consider the ratio $w=\psi^D/\psi$ in the latter
case. Since any different choice of linearly independent solutions
of the SE corresponds to a M\"obius transformation of $w$, we can
choose
$$
\psi^D_{\;\stackrel{\sim}{q\rightarrow-\infty}\;}a_-e^{P_-q}\
,\qquad\qquad
\psi^D_{\;\stackrel{\sim}{q\rightarrow+\infty}\;}a_+e^{P_+q}\ ,
$$
and
$$
\psi_{\;\stackrel{\sim}{q\rightarrow-\infty}\;}b_-e^{-P_-q}\
,\qquad\qquad
\psi_{\;\stackrel{\sim}{q\rightarrow+\infty}\;}b_+e^{-P_+q}\ ,
$$
were by $\sim$ we mean that $\psi^D$ and $\psi$ either diverge or
vanish ``at least as". Their ratio has the asymptotic
$$
{\psi^D\over\psi}{}_{\;\stackrel{\sim}{q\rightarrow-\infty}\;}c_-e^{2P_-q}
\rightarrow0\ ,\quad{\psi^D\over\psi}{}_{\;\stackrel{\sim}{q
\rightarrow+\infty}\;}c_+e^{2P_+q}\rightarrow\pm\infty\ ,
$$
so that $w$ cannot satisfy Eq.(\ref{specificandoccnn}). This
concludes the alternative proof of the fact that, in the case
(\ref{asintoticopiumenofgt}), the existence of the $L^2(\RR)$
solution is a necessary condition in order (\ref{specificandoccnn})
be satisfied. The fact that this is a sufficient condition has been
proved previously in deriving Eq.(\ref{arbitygvxy}).

The above results imply that the usual quantized spectrum arises as
a consequence of the EP.
Let us note that we are considering real solutions of the SE. Thus,
apparently, in requiring the existence of an $L^2(\RR)$ solution,
one should specify the existence of a real $L^2(\RR)$ solution.
However, if there is an $L^2(\RR)$ solution $\psi$, this is unique
up to a constant, and since also $\bar\psi\in L^2(\RR)$ solves the
SE, we have that an $L^2(\RR)$ solution of the SE is real up to a
phase.

In the present investigation we elaborated on two main theorems that underlie
the formulation of the quantum Hamilton-Jacobi equation from an equivalence
postulate. One
should regard this postulate as providing a novel starting point
for formulating quantum mechanics. It also provides an arena to reexamine
many of the tenants of the conventional approaches, and in this regard we note
the related work of Floyd \cite{Floyd}.
There are other topics in the derivation of the quantum Hamilton-Jacobi
equation from the
EP we did not consider here. These include dualities
and geometrical structures which also appeared in other investigations
\cite{vari,reviews,140,seealso,Appleby:1999vh}.

\bibliography{apssamp}

\begin{thebibliography}{00}



\bibitem{1}
A.~E.~Faraggi and M.~Matone,
Phys.\ Lett.\ B {\bf 450}, 34 (1999);
Phys.\ Lett.\ B {\bf 437}, 369 (1998);
Phys.\ Lett.\ A {\bf 249}, 180 (1998);
Phys.\ Lett.\ B {\bf 445}, 77 (1998);
%
Phys.\ Lett.\ B {\bf 445}, 357 (1999);
Int.\ J.\ Mod.\ Phys.\ A {\bf 15}, 1869 (2000).

\bibitem{BFM}
G.~Bertoldi, A.~E.~Faraggi and M.~Matone,
%
Class.\ Quant.\ Grav.\  {\bf 17}, 3965 (2000).

\bibitem{2}
M.~Matone,
Found.\ Phys.\ Lett.\  {\bf 15}, 311 (2002);
hep-th/0212260;
  Braz.\ J.\ Phys.\  {\bf 35}, 316 (2005).

\bibitem{mshsm}  A.~E.~Faraggi, D.~V.~Nanopoulos and K.~j.~Yuan,
  Nucl.\ Phys.\  B {\bf 335}, 347 (1990).
G.~B.~Cleaver, A.~E.~Faraggi and D.~V.~Nanopoulos,
  Phys.\ Lett.\  B {\bf 455}, 135 (1999).
 B.~Assel, K.~Christodoulides, A.~E.~Faraggi, C.~Kounnas and J.~Rizos,
  arXiv:0910.3697 [hep-th].

\bibitem{Matone:1993tj}
M.~Matone,
Int.\ J.\ Mod.\ Phys.\ A {\bf 10}, 289 (1995).

\bibitem{Ovsienko1}
V.~Ovsienko, ``Lagrange Schwarzian derivative and symplectic Sturm
theory,'' CPT-93-P-2890.
\bibitem{DMS} P.~Di~Francesco, P.~Mathieu and D.~S\'en\'echal, ``Conformal Field Theory", Springer, 1996.
\bibitem{Messiah} A.~Messiah, {\it Quantum Mechanics}, Vol. 1, North--Holland, 1961.
\bibitem{Floyd}
E.~R.~Floyd, Phys.\ Rev.\  {\bf D25}, 1547 (1982);
Phys.\ Rev.\ D {\bf 26}, 1339 (1982);
Phys.\ Rev.\ D {\bf 29}, 1842 (1984);
Phys.\ Rev.\  {\bf D34}, 3246 (1986);
Int.\ J.\ Theor.\ Phys.\  {\bf 27}, 273 (1988);
Phys.\ Lett.\  {\bf A214}, 259 (1996);
%
Found.\ Phys.\ Lett.\  {\bf 9}, 489 (1996);
%
Found.\ Phys.\ Lett.\  {\bf 13}, 235 (2000);
%
Int.\ J.\ Mod.\ Phys.\ A {\bf 14}, 1111 (1999);
%
Int.\ J.\ Mod.\ Phys.\ A {\bf 15}, 1363 (2000);
quant-ph/0302128;
quant-ph/0307090.
\bibitem{vari}
G.~Reinisch, Physica {\bf A206}, 229 (1994);
Phys.\ Rev.\  {\bf A56}, 3409 (1997).
A.~E.~Faraggi,
hep-th/0411118.
R.~Carroll, Can.\ J.\ Phys.\  {\bf 77}, 319 (1999);
quant-ph/0309023;
quant-ph/0401082;
gr-qc/0406004;
quant-ph/0403156;
gr-qc/0501045.
A.~Bouda,
Found.\ Phys.\ Lett.\  {\bf 14}, 17 (2001);
%
Int.\ J.\ Mod.\ Phys.\ A {\bf 18}, 3347 (2003).
A.~Bouda and T.~Djama,
Phys.\ Lett.\ A {\bf 285}, 27 (2001);
Phys.\ Scripta {\bf 66}, 97 (2002).
A.~Bouda and F.~Hammad,
Acta Phys.\ Slov.\  {\bf 52}, 101 (2002).
M.~R.~Brown,
%
quant-ph/0102102.
T.~Djama,
quant-ph/0111121;
%
quant-ph/0111142;
quant-ph/0201003;
quant-ph/0404098.
\bibitem{reviews}
A.~E.~Faraggi,
hep-th/9910042.
R.~Carroll, ``Quantum theory, deformation and integrability,''
Elsevier, North-Holland, 2000. North-Holland Mathematics Studies
186.
J.~M.~Delhotel,
quant-ph/0401063.
R.~E.~Wyatt, ``Quantum Dynamics with trajectories,'' Springer, 2005.

\bibitem{140}
M.~Matone,
Phys.\ Lett.\ B {\bf 357}, 342 (1995);
Phys.\ Rev.\ D {\bf 53}, 7354 (1996).
G.~Bonelli and M.~Matone,
%
Phys.\ Rev.\ Lett.\  {\bf 76}, 4107 (1996);
Phys.\ Rev.\ Lett.\  {\bf 77}, 4712 (1996).
G.~Bonelli, M.~Matone and M.~Tonin,
Phys.\ Rev.\ D {\bf 55}, 6466 (1997).
M.~Matone,
Phys.\ Rev.\ Lett.\  {\bf 78}, 1412 (1997).
A.~E.~Faraggi and M.~Matone,
Phys.\ Rev.\ Lett.\  {\bf 78}, 163 (1997).
R.~W.~Carroll,
hep-th/9607219;
hep-th/9610216;
hep-th/9702138;
Nucl.\ Phys.\ B {\bf 502}, 561 (1997);
Lect.\ Notes Phys.\  {\bf 502}, 33 (1998).
I.~V.~Vancea,
Phys.\ Lett.\ B {\bf 480}, 331 (2000);
Phys.\ Lett.\ A {\bf 321}, 155 (2004).
M.~A.~De Andrade and I.~V.~Vancea,
Phys.\ Lett.\ B {\bf 474}, 46 (2000).
M.~C.~B.~Abdalla, A.~L.~Gadelha and I.~V.~Vancea,
Phys.\ Lett.\ B {\bf 484}, 362 (2000).

\bibitem{seealso}
J.~Butterfield,
quant-ph/0210140.
J.~M.~Isidro,
Int.\ J.\ Mod.\ Phys.\ A {\bf 16}, 3853 (2001);
quant-ph/0105012;
J.\ Phys.\ A {\bf 35}, 3305 (2002);
quant-ph/0112032;
J.\ Geom.\ Phys.\  {\bf 41}, 275 (2002);
Phys.\ Lett.\ A {\bf 301}, 210 (2002);
hep-th/0204178;
Mod.\ Phys.\ Lett.\ A {\bf 18}, 1975 (2003);
hep-th/0304175;
Mod.\ Phys.\ Lett.\ A {\bf 19}, 349 (2004);
Phys.\ Lett.\ A {\bf 317}, 343 (2003);
quant-ph/0310092;
Mod.\ Phys.\ Lett.\ A {\bf 19}, 1733 (2004);
hep-th/0407161;
Mod.\ Phys.\ Lett.\ A {\bf 19}, 2339 (2004);
quant-ph/0411166.
\bibitem{Appleby:1999vh}
D.~M.~Appleby,
Found.\ Phys.\  {\bf 29}, 1863 (1999);
Phys.\ Rev.\ A {\bf 65}, 022105 (2002);
quant-ph/0308114.
B.~Poirier,
J.\ Chem.\ Phys.\  {\bf 121}, 4501 (2004).
F.~Girelli, E.~R.~Livine and D.~Oriti,
Nucl.\ Phys.\ B {\bf 708}, 411 (2005).
M.~V.~John,
Found.\ Phys.\ Lett.\  {\bf 15}, 329 (2002);
quant-ph/0102087.

\end{thebibliography}

\end{document}